\documentclass[11pt]{article}

\usepackage{times}
\usepackage{graphicx}
\usepackage{amsmath}

\title{Three dimensional structure from intensity correlations}
\author{Veit Elser \\ Department of Physics, Cornell University \\ Ithaca, NY 14853-2501 \\ USA}

\begin{document}

\maketitle

\begin{center}
\parbox{5in}{We develop the analysis of x-ray intensity correlations from dilute ensembles of identical particles in a number of ways. First, we show that the 3D particle structure can be determined if the particles can be aligned with respect to a single axis having a known angle with respect to the incident beam. Second, we clarify the phase problem in this setting and introduce a data reduction scheme that assesses the integrity of the data even before the particle reconstruction is attempted. Finally, we describe an algorithm that reconstructs intensity and particle density simultaneously, thereby making maximal use of the available constraints.
} 
\end{center}

\section{Introduction}

As first pointed out by Kam \cite{Kam}, intensity fluctuations in an x-ray experiment with ensembles of non-oriented identical particles can provide structural information about the particles themselves. Correlations in the fluctuations arise from sets of photons that are scattered from the same particle and as such effectively provide a signal from individual particles. This technique does not place demands on the spatial coherence of the source --- provided it spans at least one particle --- but requires short pulses of high fluence and was largely not developed for that reason. With the availability of high intensity free-electron laser sources in recent years the situation has changed, and there is renewed interest in developing Kam's idea into a practical method. This effort has been led by Dilano and co-workers, with theoretical contributions \cite{Saldin1, Saldin2} as well as proof-of-principle experiments \cite{Saldin3, Saldin4}.

There is currently no known method for extracting 3D structure from intensity fluctuation data. Moreover, a constraint counting argument \cite{Spencefest} shows that information derived from experiments with completely non-oriented particles is deficient for extracting 3D structure. In this paper we revive the prospects of Kam's idea in the 3D realm provided the particles in the ensemble can be partially oriented. The type of alignment required is illustrated with pyramidal particles in Figure 1. As a result of native or engineered interactions and geometry, we propose that the particles adsorb on a substrate in such a way that a particular oriented axis within the particle is aligned with the substrate normal with near 100\% probability\footnote{With considerable additional effort the method we describe might be extended to allow for multiple modes of adsorption, in other words, when the dice are not as strongly loaded to land on one particular face as considered here.}. Other means of achieving this degree of partial orientation would achieve the same goal. The concentration of their energy density makes substrates more efficient alignment mechanisms than laser fields, but this comes at the expense of introducing background and making certain ranges of the orientation angle inaccessible to the experiment. Membrane protein complexes, axially oriented in a synthetic lipid membrane, a liquid substrate, would be natural targets for this technique \cite{Saldin2}.

The fluctuations that form the basis of Kam's idea are not due to photon number fluctuations but arise from different realizations of the particle orientations. In our case, these are limited to rotations about the substrate normal. If the particles undergo rotational diffusion, then the duration of each intensity measurement must be short compared to the diffusion time scale. When diffusion is absent, the sample substrate must be translated in order that a different set of particles, with different orientations, intercepts the beam. In either case, and from a signal perspective, it is always advantageous to focus the beam on the smallest number of particles $N$, if this can be done so without adversely affecting beam divergence and the integrated fluence. That is because, while the mean intensities recorded at two positions on the detector are independent of $N$, their covariance --- the source of structure information --- is diminished by $1/N$. Even when the background signal at the two pixels is perfectly uncorrelated, in being independent of $N$ it will dominate the error in the covariance estimation for large $N$. There are other factors, however, that rule out small $N$ even when this does not sacrifice the quality of the beam. Chief among these is particle damage, which must be kept at a low rate if subsequent measurements are taken on a single set of rotationally diffusing particles. 

\begin{figure}[t]
\begin{center}\scalebox{1.2}{\includegraphics{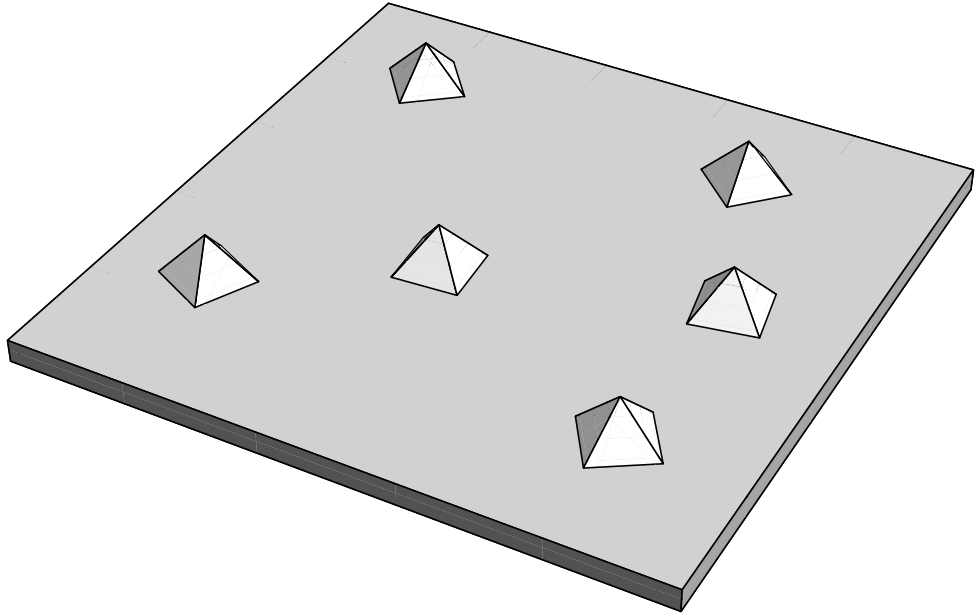}}
\end{center}
\begin{center}
\parbox{6in}{\caption{Partially oriented particle ensemble realized as dice with a strong tendency to land on one particular face. Alternatively, partial alignment might be realized in the manner of membrane proteins embedded in a lipid membrane.}}
\end{center}
\end{figure}

In a nutshell, our reconstruction method involves three stages. In the \textit{data collection stage}, intensity correlations are recorded for five continuous parameters. Four of these are the $x$ and $y$ positions of two pixels on the detector and the fifth is the tilt angle $\theta$ of the substrate. This 5D data set is reduced to a 3D data set in the \textit{data reduction stage}. The 3D data comprise complex numbers $I_m(r,z)$, or angular harmonics, giving the Fourier series expansion coefficients of the particle intensity on families of circles in a frame fixed to the particle. The data reduction also provides an internal consistency criterion, similar to principal component analysis, for establishing ranges for the radial ($r$) and longitudinal ($z$) frequency content, as well as the maximum harmonic order $M$ that can be extracted from the data. The data reduction stage leaves $M/2$ phases in the intensity reconstruction undetermined. Finally, in the \textit{particle reconstruction stage}, two phase problems are solved jointly that determine the 3D particle density from the (3D) angular harmonics. In addition to the usual phases encountered when reconstructing from intensity data, the $M/2$ unknown phases in the angular harmonics from the second stage are determined from the property that the intensity is consistent with a particle of compact support.

\section{Data collection}

The coordinate frame for data collection is defined by the incident x-ray beam, as $\mathbf{z}$-axis, and orthogonal $\mathbf{x}$ and $\mathbf{y}$ axes in the plane of the detector. With the sample located at the origin, the unscattered beam would hit the detector at $x=y=0$, $z=Z$. The axes on the detector are chosen with reference to the axis about which the substrate holding the sample is rotated. For tilt angle $\theta=0$ the substrate and detector planes are parallel and in-plane axes $\mathbf{x'}$ and $\mathbf{y'}$ on the substrate coincide with those on the detector. In all the measurements the substrate is tilted about the $\mathbf{y'}$ axis, leading to the following relationship between the lab and substrate axes:
\begin{subequations}\label{rot1}
\begin{align}
\mathbf{x}&=\cos{\theta}\, \mathbf{x'}+\sin{\theta}\, \mathbf{z'}\\
\mathbf{y}&=\mathbf{y'}\\
\mathbf{z}&=-\sin{\theta}\, \mathbf{x'}+\cos{\theta}\, \mathbf{x'}.
\end{align}
\end{subequations}
The $\mathbf{z'}$ axis is the substrate normal and coincides with an alignment axis fixed in the body of each particle adsorbed on the substrate. A third set of axes, fixed in a particular particle, is free to rotate with respect to the substrate about its $\mathbf{z'}$ axis by some angle $\alpha$:
\begin{subequations}\label{rot2}
\begin{align}
\mathbf{x'}&=\cos{\alpha}\, \mathbf{x''}+\sin{\alpha}\, \mathbf{y''}\\
\mathbf{y'}&=-\sin{\alpha}\, \mathbf{x''}+\cos{\alpha}\, \mathbf{y''}\\
\mathbf{z'}&=\mathbf{z''}.
\end{align}
\end{subequations}
In the experiment there is no control over the angle $\alpha$ and we assume it takes uniformly distributed random values for each particle.

We can now express the photon wave vector change, seen in the laboratory, in the frame fixed to a particular particle. For intensity correlation measurements we are interested in the intensities at two points on the detector plane, that we label 1 and 2. The source of the scattered photons is the same particle, since the covariance vanishes for different particles whose rotation angles $\alpha$ are assumed to be independent. Using equations \eqref{rot1} and \eqref{rot2}, the wave vector transferred to photons detected at coordinates $x_1$ and $y_1$ on the detector is
\begin{align}
\mathbf{q}_1&=\left(\frac{2\pi}{\lambda Z}\right)\mathbf{\tilde{q}}_1\\
\mathbf{\tilde{q}}_1&=\left(x_1\cos{\theta}\cos{\alpha}-y_1\sin{\alpha}\right)\mathbf{x''}+\left(x_1\cos{\theta}\sin{\alpha}+y_1\cos{\alpha}\right)\mathbf{y''}+\left(x_1\sin{\theta}\right)\mathbf{z''}\label{q1}
\end{align}
where $\lambda$ is the wavelength and there is a similar equation for $\mathbf{q}_2$. In subsequent equations we will express all wave vectors, as here, in units of $2\pi/(\lambda Z)$. In \eqref{q1} we made the small angle approximation, neglecting corrections of order $(x/Z)^2$ and $(y/Z)^2$.

For both pixels in our correlation measurement we define coordinates $r$, $\phi$ and $z$ as follows:
\begin{subequations}\label{rphiz}
\begin{align}
x_1\cos{\theta}&=r_1\sin{\phi_1}\label{rphiz1}\\
y_1&=r_1\cos{\phi_1}\label{rphiz2}\\
x_1\sin{\theta}&=z_1,\label{rphiz3}
\end{align}
\end{subequations}
and a similar set of definitions for pixel 2. Let $I(x,y;\theta)$ be the (time integrated) intensity recorded in one measurement at detector pixel coordinates $x$ and $y$ and tilt angle $\theta$. In the experiment we measure the following correlation function:
\begin{equation}\label{corr}
C(x_1,y_1;x_2,y_2;\theta)=\mathrm{cov}\left(I(x_1,y_1;\theta),I(x_2,y_2;\theta)\right).
\end{equation}
The covariance is estimated in the usual way, from averages of intensities and their products taken from a set of diffraction patterns at fixed tilt. Let $I_\mathrm{p}(x,y,z)$ be the intensity (squared Fourier magnitude) of the particle density as a function of the scaled wave vector components in the body-fixed coordinate system. Using \eqref{q1} and the definitions \eqref{rphiz}, the covariance is expressed in terms of the particle intensity as
\begin{align}\label{cov}
\mathrm{cov}&\left(I(x_1,y_1;\theta),I(x_2,y_2;\theta)\right)/N(\theta)=\\
&\langle\, I_\mathrm{p}\left(r_1\sin{(\phi_1-\alpha)},r_1\cos{(\phi_1-\alpha)},z_1\right)\,I_\mathrm{p}\left(r_2\sin{(\phi_2-\alpha)},r_2\cos{(\phi_2-\alpha)},z_2\right)\,\rangle_\alpha\nonumber\\
&-\langle\, I_\mathrm{p}\left(r_1\sin{(\phi_1-\alpha)},r_1\cos{(\phi_1-\alpha)},z_1\right)\,\rangle_\alpha\,
\langle\, I_\mathrm{p}\left(r_2\sin{(\phi_2-\alpha)},r_2\cos{(\phi_2-\alpha)},z_2\right)\,\rangle_\alpha,\nonumber
\end{align}
where $\langle\, \cdots\rangle_\alpha$ is an average over the uniformly distributed particle rotation angle $\alpha$ and $N(\theta)$ is the number of particles intercepted by the incident beam. In this equation the definition of the particle intensity $I_\mathrm{p}$ has absorbed several factors, including beam fluence, that we assume are kept constant. In the following we further assume that the particle number density on the substrate is constant so that (again omitting a constant factor) we have
\begin{equation}
N(\theta)=1/\cos{\theta}.
\end{equation}

Rotating both the tilted substrate and the detector by $\pi$ about the beam axis should not change the correlations. This implies the following symmetry property of the correlation function:
\begin{equation}\label{sym}
C(x_1,y_1;x_2,y_2;\theta)=C(-x_1,-y_1;-x_2,-y_2;-\theta).
\end{equation}
It is therefore sufficient to only collect data for positive tilt angles.

\section{Data reduction}

As a first step in reducing the data we represent the particle density in terms of its angular harmonics with respect to rotations about the body axis parallel to the substrate normal:
\begin{equation}
I_\mathrm{p}(r \sin{\varphi},r\cos{\varphi},z)=\sum_m e^{i m\varphi} I_m(r,z).
\end{equation}
Using this definition in \eqref{cov} and the reality property of $I_\mathrm{p}$ we can evaluate the average over $\alpha$ and obtain
\begin{equation}\label{coupled}
C(x_1,y_1;x_2,y_2;\theta) \cos{\theta}=\sum_{m\ne 0} e^{i m(\phi_1-\phi_2)}I_m(r_1,z_1)I_m^*(r_2,z_2).
\end{equation}
We can decouple these equations by averaging both sides of \eqref{coupled} against the factor $e^{-i m'(\phi_1-\phi_2)}$ in such a way that $\phi_1-\phi_2$ is uniformly sampled on an interval of $2\pi$ radians. Moreover, the decoupling is especially simple if we can do this while keeping the parameters $r_1$, $z_1$, $r_2$ and $z_2$ all constant.

The desired decoupling is achieved by a one-parameter average for which we use either $\phi_1$ or $\phi_2$, depending on the value of
\begin{equation}
A=\frac{r_1 z_2}{r_2 z_1}.
\end{equation}
When $|A|<1$ we use $\phi_1$ as the parameter; otherwise, by switching the labels 1 and 2 we restore $|A|<1$ and the parameter becomes $\phi_2$. In the following we assume the labels are such that $|A|<1$, making $\phi_1$ the averaging parameter. Taking the ratio of equations \eqref{rphiz1} and \eqref{rphiz3} for both labels we obtain an equation for the second azimuthal angle in terms of the first,
\begin{equation}\label{phi2}
\phi_2(\phi_1)=\arcsin{(A \sin{\phi_1})},
\end{equation}
where we take the branch that is continuous with $\phi_2\to 0$ in the limit $A\to 0$. In the opposite limit, $A\to \pm 1$, $\phi_2$ is a sawtooth of amplitude $\pi/2$. There is a second branch $\pi-\phi_2(\phi_1)$ that we discuss below. For all $A$ (in the range $|A|<1$), the function
\begin{equation}
\varphi(\phi_1)=\phi_1-\phi_2(\phi_1)
\end{equation}
is monotonic and ranges over exactly $2\pi$ radians whenever $\phi_1$ does. Since we perform the average with respect to the parameter $\phi_1$, we need to include the Jacobian
\begin{equation}
J(\phi_1)=\left|\frac{d\varphi}{d\phi_1}\right|=1-\frac{A \cos{\phi_1}}{\sqrt{1-A^2\sin^2{\phi_1}}}
\end{equation}
in order that $\varphi(\phi_1)$ is uniformly sampled.

Equations \eqref{rphiz} and the companion equations involving label 2 determine how the detector parameters $x_1$, $y_1$, $x_2$, $y_2$ and $\theta$ need to be varied along with $\phi_1$ and $\phi_2(\phi_1)$ so that the body-fixed parameters $r_1$, $z_1$, $r_2$ and $z_2$ are kept constant. Comparing \eqref{rphiz1} and \eqref{rphiz3} (and their companions) we see that the tilt angle must satisfy
\begin{equation}\label{theta}
\arctan{\left(\frac{z_1}{r_1\sin{\phi_1}}\right)}=\theta(\phi_1)=\arctan{\left(\frac{z_2}{r_2\sin{\phi_2(\phi_1)}}\right)},
\end{equation}
where compatibility follows from our definition \eqref{phi2} of $\phi_2(\phi_1)$. We take the branch of $\theta(\phi_1)$ that lies in the range $|\theta|<\pi/2$, the center of which ($\theta=0$) corresponds to normal beam incidence on the substrate. Formulas for the other detector parameters are summarized below:
\begin{subequations}\label{xy12}
\begin{align}
x_1(\phi_1)&=\frac{z_1}{\sin{\theta(\phi_1)}}\\
x_2(\phi_1)&=\frac{z_2}{\sin{\theta(\phi_1)}}\\
y_1(\phi_1)&=r_1\cos{\phi_1}\\
y_2(\phi_1)&=r_2\cos{\phi_2(\phi_1)}.
\end{align}
\end{subequations}

Performing the one-parameter average of the correlation function we arrive at the decoupled equations:
\begin{subequations}\label{Cm}
\begin{align}
C_m(r_1,z_1;r_2,z_2)&=\int_0^{2\pi}\frac{d\phi_1}{2\pi}J(\phi_1) e^{-i m\varphi(\phi_1)}\cos{\theta(\phi_1)}C(x_1(\phi_1),y_1(\phi_1);x_2(\phi_1),y_2(\phi_1);\theta(\phi_1))\label{Cm1}\\
&=I_m(r_1,z_1)I_m^*(r_2,z_2).\label{Cm2}
\end{align}
\end{subequations}
Before we study equation \eqref{Cm} in greater depth we examine the second branch of the solution to \eqref{phi2} given by
\begin{equation}
\phi'_2(\phi_1)=\pi-\phi_2(\phi_1).
\end{equation}
Switching to the new parameter
\begin{equation}
\phi'_1=\phi_1+\pi,
\end{equation}
we find that the previously defined functions are changed in simple ways. Starting with
\begin{equation}
\phi_2(\phi_1)=-\phi_2(\phi'_1)
\end{equation}
we find
\begin{align}
\varphi'(\phi_1)&=\phi_1-\phi'_2(\phi_1) = \phi'_1-\phi_2(\phi'_1)-2\pi\\
&=\varphi(\phi'_1)-2\pi
\end{align}
so that (up to an irrelevant $2\pi$ shift) the averaging phase for the second branch is the same as for the first branch when expressed in terms of the parameter $\phi'_1$. The same is true of the Jacobian:
\begin{equation}
J'(\phi_1)=J(\phi'_1).
\end{equation}
Finally, using \eqref{theta} and \eqref{xy12} one easily verifies the tilt and detector parameters for the second branch simply undergo a reversal of sign:
\begin{subequations}
\begin{align}
\theta'(\phi_1)&=-\theta(\phi'_1)\\
x'_1(\phi_1)&=-x_1(\phi'_1)\\
x'_2(\phi_1)&=-x_2(\phi'_1)\\
y'_1(\phi_1)&=-y_1(\phi'_1)\\
y'_2(\phi_1)&=-y_2(\phi'_1)
\end{align}
\end{subequations}
The correlation average for the second branch is thus identical to \eqref{Cm1} with all the arguments of the correlation function replaced by their negatives. 
The second branch therefore provides no additional information once symmetry \eqref{sym} is taken into account.

In equation \eqref{Cm1} we average the correlations $C$, between various pairs of pixels on the detector and various tilts, for particular harmonics $m$ and define a correlation $C_m$ intrinsic to the particle in that its arguments are spatial frequencies $r$ and $z$ in the body-fixed frame. The two pixels move on circular arcs in the average, since equations \eqref{rphiz} and their companions for pixel 2 imply
\begin{subequations}\label{arcs}
\begin{align}
x_1^2+y_1^2&=r_1^2+z_1^2\\
x_2^2+y_2^2&=r_2^2+z_2^2,
\end{align}
\end{subequations}
where the right sides are constant in any average.
The left sides in these equations represent the magnitudes of the wave vector changes seen in the laboratory while the right sides are the magnitudes of the spatial frequencies in the particle responsible for those changes. Figure 2 shows the orbits of two pixels and the corresponding tilt in an average for one particular choice of $r_1$, $z_1$, $r_2$ and $z_2$. Since all orbits visit tilt angle $\theta=\pi/2$, where the beam is parallel to the substrate, the correlation function needs to be interpolated over the inaccessible range before the average \eqref{Cm} is evaluated. By using the symmetry \eqref{sym} it is only necessary to acquire data for $\theta$ between $0$ and $\pi/2$.

\begin{figure}[t]
\begin{center}\scalebox{1.3}{\includegraphics{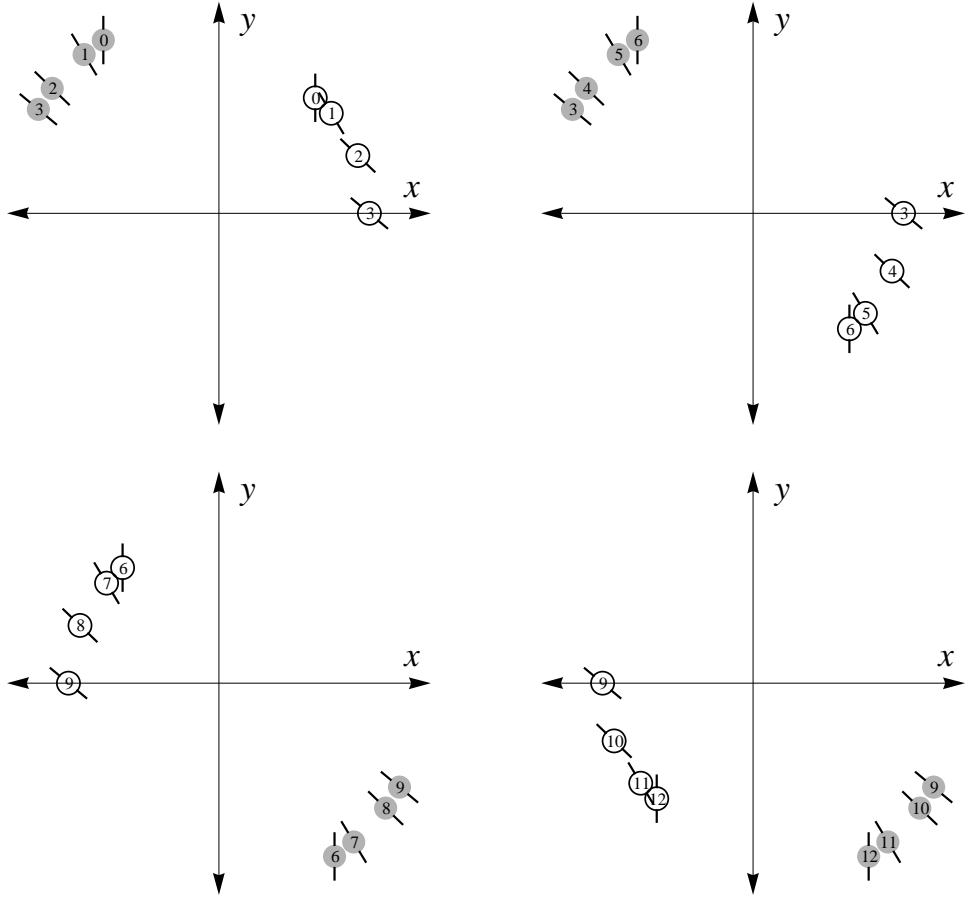}}
\end{center}
\begin{center}
\parbox{6in}{\caption{Pairs of detector pixels (labeled 0-12) in one averaging orbit \eqref{Cm}  for the case $r_1=6$, $z_1=5$, $r_2=9$ and $z_2=-6$. Lines through the circles show the tilt angle $\theta$, where we have used symmetry \eqref{sym} to keep the range of tilts between 0 and $\pi/2$. The tilt is smallest at pairs 3 and 9 and reaches $\pi/2$ (grazing incidence) at pairs 0, 6, 12.}}
\end{center}
\end{figure}

In the final step of data reduction the angular harmonic amplitudes $I_m(r,z)$ are extracted from the averaged correlation function $C_m$. For fixed $m$, the function $C_m(r_1,z_1;r_2,z_2)$ is a Hermitian matrix with row indices $r_1,z_1$ and column indices $r_2,z_2$. The content of \eqref{Cm2} is the fact that this matrix has rank one and its unique eigenvector with non-zero eigenvalue, up to an overall phase, is $I_m(r,z)$.

The procedure for extracting $I_m(r,z)$ is similar to principal component analysis and tests the integrity of the data prior to the phase reconstructions that follow. First a set of radial ($r$) and longitudinal ($z$) spatial frequency samples are selected that, by \eqref{arcs}, lie within the beamstop and edge of the detector. The density of the samples is set by the speckle scale of the single particle diffraction pattern. Next, the matrices $C_m$ are calculated by averaging correlations for the chosen $r$, $z$ samples according to \eqref{Cm1} and for $|m|\le M/2$, where the maximum harmonic $M$ is set by the angular speckle scale at the edge of the detector. For each matrix $C_m$ one then obtains the dominant normalized eigenvector $V_m(r,z)$ and corresponding eigenvalue $\lambda_m$ and forms the estimates
\begin{equation}\label{eigen}
I_m(r,z)=\sqrt{\lambda_m}\, V_m(r,z).
\end{equation}
From the relative magnitudes of the subdominant eigenvalues of $C_m$ one can assess the quality of the data at the chosen resolution. If these are significant, then it may be necessary to work with lower resolution blocks of $C_m$, reduce $M$, or fix the interpolation of the correlation function at $\theta=\pi/2$ until single-eigenvector dominance is realized for the $C_m$. On the other hand, if no amount of data truncation satisfies this condition then the data are too flawed, \textit{e.g.} by background, to proceed any further.

We use the ratio of the largest eigenvalue magnitude to the sum of the magnitudes of all the eigenvalues of $C_m$ as an internal measure of the consistency of the reduced data,
\begin{equation}\label{sigma}
\sigma_m=\frac{\|C_m\|_\infty}{\|C_m\|_1},
\end{equation}
which, as expressed by the notation, is the ratio of the spectral and trace norms of the matrix $C_m$. In the absence of noise and other complications $\sigma_m=1$.

Since the phases of the eigenvectors $V_m$ in \eqref{eigen} are undetermined, our information about the intensity is incomplete at the level of $M/2$ unknown phase angles. These angles have to be reconstructed on the basis of additional constraints before the 3D intensity of the particle can be synthesized from its angular harmonics $I_m(r,z)$. Since $M$ scales only with the diameter of the particle, and not its volume, this missing information is relatively minor in comparison with the many phases that must be reconstructed to obtain the particle density from the intensity. An algorithm for reconstructing both types of phases is described in section \ref{partrecon}.

In contrast to the $m\ne 0$ harmonics $I_m(r,z)$ that require correlation analysis, the $m=0$ harmonic is obtained by a straightforward average of the intensity and is therefore much easier to acquire. There is also no phase ambiguity in $I_0(r,z)$. Instead of \eqref{corr} we define
\begin{equation}
A(x,y;\theta)=\mathrm{ave}\left(I(x,y;\theta)\right)
\end{equation}
and \eqref{cov} is replaced by
\begin{subequations}
\begin{align}
\mathrm{ave}\left(I(x,y;\theta)\right)/N(\theta)&=
\langle\, I_\mathrm{p}\left(r\sin{(\phi-\alpha)},r\cos{(\phi-\alpha)},z\right)\,\rangle_\alpha\\
&=I_0(r,z).
\end{align}
\end{subequations}
We can aggregate the three parameter data on the left in a way that tests the two parameter model on the right. One method is to use the earlier parameterization \eqref{theta} and \eqref{xy12} to define harmonics
\begin{equation}\label{powderave}
A_m(r,z)=\int_0^{2\pi}\frac{d\phi}{2\pi}e^{-i m \phi} \cos{\theta(\phi)}\, A\left(x(\phi),y(\phi);\theta(\phi)\right)
\end{equation}
such that $A_0(r,z)$ becomes the estimate of $I_0(r,z)$ and power in the nonzero harmonics indicates data inconsistency.

\subsection{Data reduction for axial projections}

The problem of reconstructing an axial projection from data at tilt angle $\theta=0$ has been studied extensively by Dilano and co-workers \cite{Saldin2, Saldin3}. In this section we consider this special case and recover the main result of the previous studies. The final step of our data reduction, however, is different and is the starting point for the improved phasing method described in section \ref{reconaxial}.

By equations \eqref{rphiz} the body-frame wave vector $z$ vanishes at tilt angle $\theta=0$. The corresponding limit of the particle intensity harmonic function $I_m(r,0)$ 
provides information about the projected particle. Equation \eqref{coupled} reduces to
\begin{equation}\label{coupledaxial}
C(x_1,y_1;x_2,y_2;0) =\sum_{m\ne 0} e^{i m(\phi_1-\phi_2)}I_m(r_1,0)I_m^*(r_2,0).
\end{equation}
where the pixel coordinates are now simply
\begin{subequations}
\begin{align}
x_1(\phi_1)&= r_1 \sin{\phi_1}\\
y_1(\phi_1)&= r_1 \cos{\phi_1},
\end{align}
\end{subequations}
and similarly for pixel 2. Specifying the angle of pixel 2 as $\phi_2=\phi_1-\varphi$, the right side of \eqref{coupledaxial} is independent of $\phi_1$ and we may average the data on the left side with respect to this angle. Decoupling with respect to the harmonic $m$ is now accomplished by a simple integral with respect to $\varphi$:
\begin{subequations}
\begin{align}
C_m(r_1,0\,;r_2,0)&=\int_0^{2\pi}\frac{d\varphi}{2\pi}e^{-i m\varphi}\int_0^{2\pi}\frac{d\phi_1}{2\pi} C(x_1(\phi_1),y_1(\phi_1);x_2(\phi_1-\varphi),y_2(\phi_1-\varphi);0)\label{Cmaxial1}\\
&=I_m(r_1,0)I_m^*(r_2,0).\label{Cmaxial2}
\end{align}
\end{subequations}

\begin{figure}[t]
\centering
\includegraphics[width=4.in]{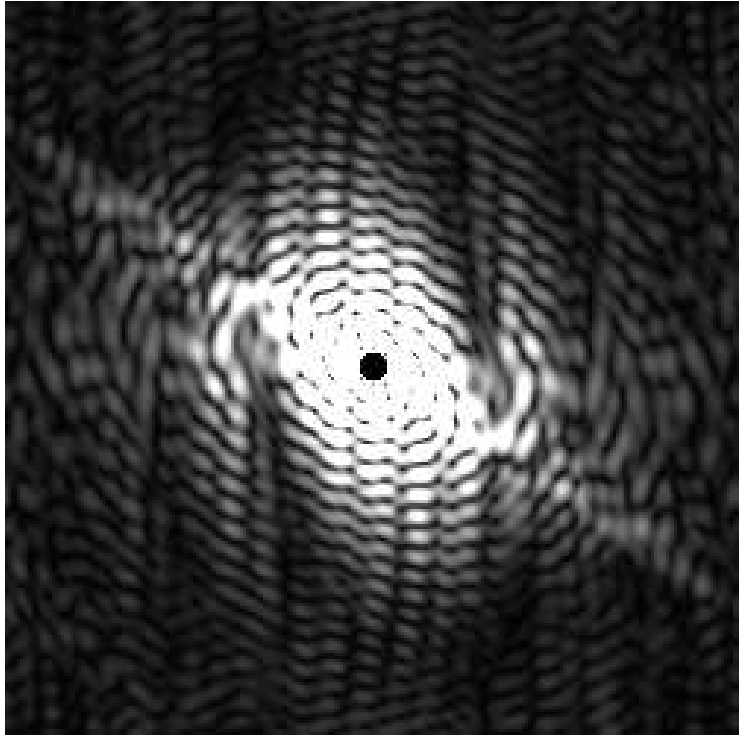}
\parbox{6in}{\caption{Single-particle intensity $I_\mathrm{p}(x,y,0)$ of the test particle used in the reconstruction demonstrations. The dark circle at the center corresponds to the missing beam stop data.}}
\end{figure}

As in the general case, \eqref{Cmaxial2} implies that the intensity harmonics are given by
\begin{equation}
I_m(r,0)=\sqrt{\lambda_m}V_m(r),
\end{equation}
where $V_m(r)$ is, up to a phase, the unique eigenvector of the Hermitian matrix $C_m(r_1,0\,;r_2,0)$ with non-zero eigenvalue $\lambda_m$. The integrity of the data is assessed using the ratio $\sigma_m$ \eqref{sigma} as in the general case. 
The harmonic $I_0(r,0)$ is just the $z=\theta=0$ limit of \eqref{powderave} and corresponds, for $m=0$, to a simple angular (powder) average of the intensity.

\begin{figure}[t]
\centering
\includegraphics[width=4.in]{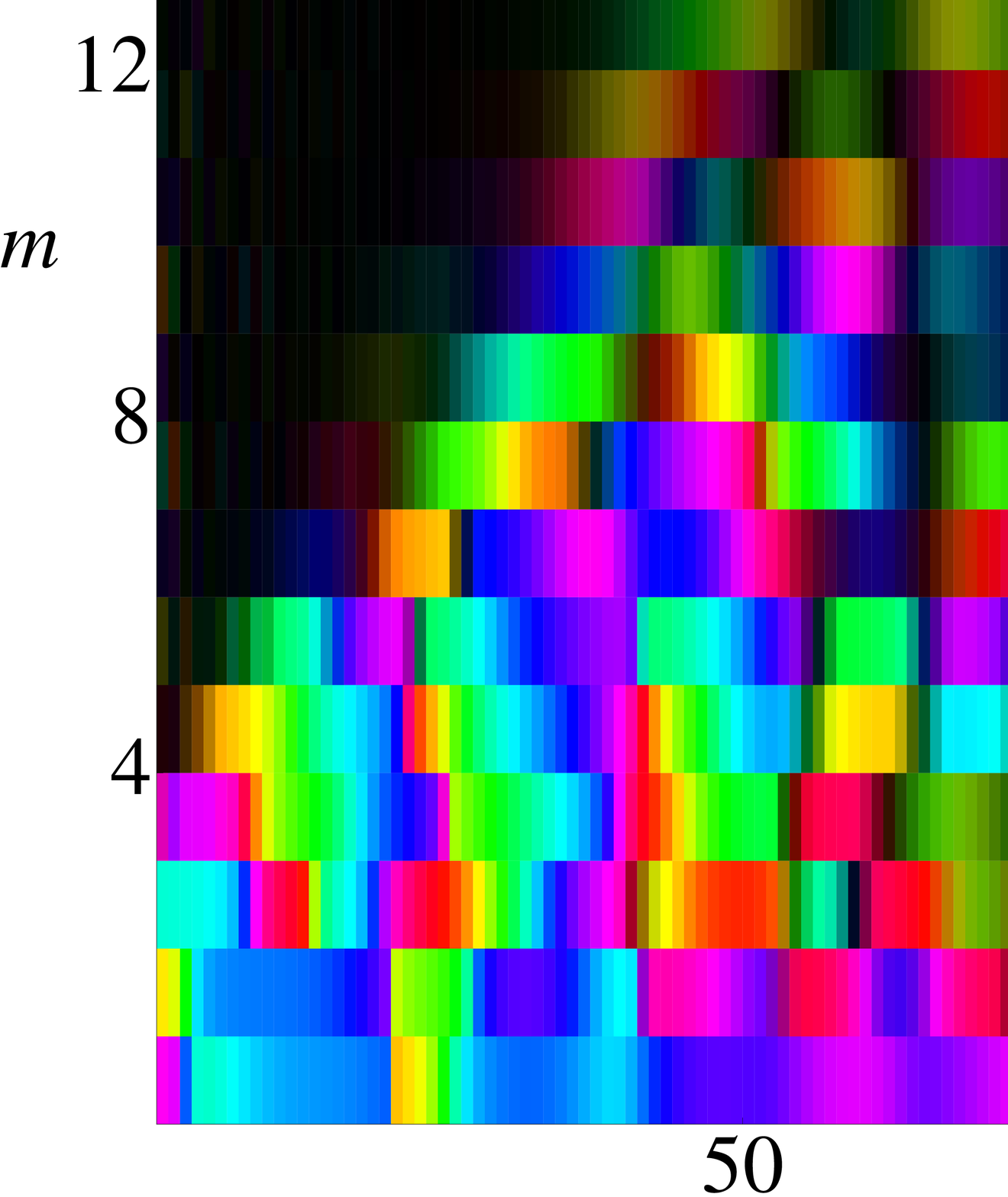}
\parbox{6in}{\caption{Result of the data reduction stage in the reconstruction of the axial projection of a test particle density. The plot shows the complex angular harmonics $I_m(r,0)$ derived from the principal eigenvectors of the angularly averaged correlation matrix \eqref{Cmaxial1}, for each $m$, of the intensity in Figure 3. Magnitude and phase are rendered as brightness and hue, respectively. There is a global phase ambiguity for each $m$ (row), the angles $\alpha_m$, that the data reduction does not determine.}}
\end{figure}

We illustrate data reduction for the axial projection of a simple test particle whose intensity is shown in Figure 3. Uncorrelated, uniform amplitude and normally distributed noise $\eta$ was added to the true intensity correlations $C$ to simulate the effects of background. Our signal-to-noise ratio in this model is defined by
\begin{equation}
\mathrm{SN}=C_\mathrm{rms}/\eta_\mathrm{rms},
\end{equation}
where the root-mean-square amplitude of $C$ is evaluated as a uniform average over all the pairs of measured pixels.

After performing the angular averages \eqref{Cmaxial1} we extract the angular harmonics $I_m(r,0)$ as the dominant eigenvectors of the matrices $C_m$. The result of this for the test particle data with zero noise is shown in Figure 4. That the harmonics $I_m(r,0)$ have a lower triangular support is a direct consequence of the uniform size of speckles in the intensity $I_\mathrm{p}$. Because our algorithm for extracting normalized eigenvectors used an unspecified convention for the phase, there is no significance to the global phase of any of the rows of the harmonics plotted in Figure 4 (an arbitrary color-wheel rotation may be applied to each one). The true phases $\alpha_m$ have to be determined in the final particle reconstruction stage.

When noise is added to the intensity correlations the harmonics $I_m(r,0)$ shown in Figure 4 become noisy as well. This is well diagnosed quantitatively through the principal eigenvector dominance measure $\sigma_m$ and is shown plotted for our test particle in Figure 5 for three noise amplitudes. We will see that for our particular test particle as few as 10 reliably extracted harmonics ($\mathrm{SN}=200$) is nearly sufficient for density reconstruction while less than 5 harmonics ($\mathrm{SN}=5$) is insufficient.

\begin{figure}[t]
\centering
\includegraphics[width=5.in]{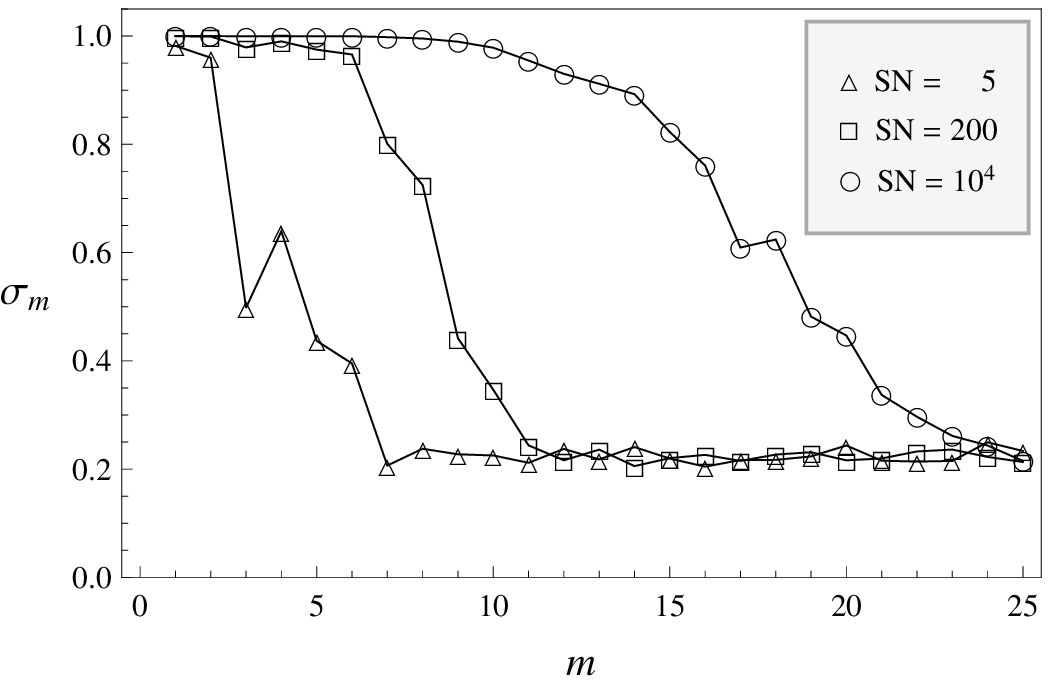}
\parbox{6in}{\caption{Dominance of the principal eigenvector as measured by the quantity $\sigma_m$ defined in \eqref{sigma} from the test particle data reduction and three level of noise.}}
\end{figure}

\section{Particle reconstruction}\label{partrecon}

As remarked in connection with equation \eqref{eigen}, additional constraints are required to determine the $M/2$ phases of the intensity angular harmonics $I_m(r,z)$. Non-negativity of the intensity is clearly one constraint that can be used. This constraint, however, is weak in comparison with the much stronger constraint that the intensity is the squared Fourier magnitude of a particle density having compact support. And since the particle density is our actual goal, we should approach the reconstruction of the $M/2$ intensity phases as part of a single process applied to intensity and density jointly. As we will show, the joint reconstruction is a relatively straightforward application of a general procedure once the constraints and the projections to them are set down \cite{diffMap}.

There are three constraints that apply to the pair $\{\hat{\rho},I\}$, where $\hat{\rho}$ is the Fourier transform of the particle density $\rho$ and we have dropped the subscript $\mathrm{p}$ on the particle intensity $I$:
\begin{subequations}
\begin{align}
\mathrm{support: }&\qquad\mathrm{supp}(\rho)\subset S,\label{con1}\\
\mathrm{data: }&\qquad{\textstyle \int_0^{2\pi}\frac{d\varphi}{2\pi}} e^{-i m\varphi}I(r \sin{\varphi},r \cos{\varphi},z)=e^{i \alpha_m}I_m(r,z),\label{con2}\\
\mathrm{compatibility: }&\qquad |\hat{\rho}|^2=I.\label{con3}
\end{align}
\end{subequations}
The first two apply to $\hat{\rho}$ and $I$ individually and can therefore be combined into a single constraint projection. Projecting $\hat{\rho}$ to the support constraint is no different from how this is done in the standard phase reconstruction problem. First $\hat{\rho}$ is Fourier transformed, the resulting density $\rho$ is set to zero outside the support region $S$ and also in its interior when the density is negative, and the result of this is inverse Fourier transformed to give the projection $\hat{\rho}'$. The data constraint involves the harmonics $I_m(r,z)$ given by the data reduction and the unknown phases $\alpha_m$. Details of the corresponding constraint projection, which in effect determines the phases $\alpha_m$, are given below.

With the support and data constraints combined, the compatibility constraint becomes the second constraint in the general scheme where a problem is divided into just two easy constraints. This division of the reconstruction problem is similar to the strategy known as \textit{divide and concur} \cite{DC}. The variables $\hat{\rho}$ and $I$ are to a large extent redundant representations of the same information and by making them independent we facilitate the projection to the ``divided" constraints (support and data). In our case ``concurrence" of the redundant variables is imposed by the compatibility constraint, which takes a nonlinear form.

With any projection to a particular constraint set comes the understanding of the distance in the space of variables that the projection minimizes. In our case the variables are of two types: complex Fourier amplitudes $\hat{\rho}(x,y,z)$ and non-negative intensities $I(x,y,z)$, both sampled on the same Cartesian grid. The Euclidean distance in this space of variables, between reconstructions $\{\hat{\rho},I\}$ and $\{\hat{\rho}',I'\}$, should be written
\begin{equation}\label{metric1}
\Delta^2= \sum_{x,y,z}|\hat{\rho}'(x,y,z)-\hat{\rho}(x,y,z)|^2+\frac{1}{w}\sum_{x,y,z}|I'(x,y,z)-I(x,y,z)|^2,
\end{equation}
where the parameter $w$ addresses the fact that $\hat{\rho}$ and $I$ first of all have different units. Since the compatibility constraint satisfied by the projected variables, at each grid point $(x,y,z)$, is the equation
\begin{equation}\label{compatibility}
|\hat{\rho}'(x,y,z)|^2=I'(x,y,z),
\end{equation}
we see that $w$ has the same units as $I$. Although the precise value of $w$ has a direct effect on the projection $\{\hat{\rho},I\}\to \{\hat{\rho}',I'\}$, the projected values will always satisfy \eqref{compatibility}, which is independent of $w$. The performance of the reconstruction algorithm, on the other hand, can be adversely affected when $w$ is either very small or very large. In those limits one of the variable types, $\hat{\rho}$ or $I$, will be much more compliant than the other and as a result the constraint associated with it (support or data) will be explored more rapidly than the constraint imposed on its stiffer counterpart. We obtained good results in the numerical experiments described in section \ref{axialrecon} when $w$ was set to the maximum measured value of $I_0(r,z)$, near the minimum of $q^2=r^2+z^2$, and making $w$ larger or smaller than this by a factor of 2 had little effect on convergence. If necessary $w$ can be allowed to vary with $q$ while only slightly complicating the projection to the data constraint, the only other projection dependent on $w$.

Projecting to the compatibility constraint \eqref{compatibility} reduces to a one-parameter numerical optimization at each grid point $(x,y,z)$. Given input variables $\hat{\rho}(x,y,z)=v e^{i t}$ and $I(x,y,z)=I$, the output of the projection is given by $\hat{\rho}'(x,y,z)=v' e^{i t}$ and $I'(x,y,z)=I'$ where $v'$ and $I'$ satisfy
\begin{equation}
v'^2=I'
\end{equation}
and minimize
\begin{equation}
(v'-v)^2+(I'-I)^2/w.
\end{equation}
The resulting cubic equation for $v'$ always has a unique, positive solution which can be tabulated in advance for efficiency.

The nontrivial part of projecting to the data constraint involves only those spatial frequency combinations $(r,z)$ whose magnitude $q$ lies between $q_\mathrm{min}$ at the beamstop and $q_\mathrm{max}$ at the edge of the detector. For $q<q_\mathrm{min}$, the projection merely copies the input intensities into the output intensities, while for $q>q_\mathrm{max}$ (in the corners of the intensity grid) the intensities are set to zero.

We uniformly sample the half-annulus defined by $q_\mathrm{min}<q<q_\mathrm{max}$ and $r>0$ on a Cartesian grid with equal spacings for $r$ and $z$, and finer than the speckle scale. Every sample $(r_i,z_i)$ defines a circle of radius $r_i$ in a plane at level $z_i$ of the intensity. The first step of the (non-trivial) data projection is extracting the angular harmonics of the input intensity by Fourier transforming the intensity on these circles:
\begin{equation}
I_{mi}=\frac{1}{M}\sum_{j=0}^{M-1}e^{-i m\varphi_j}I(r_i \sin{\varphi_j},r_i \cos{\varphi_j},z_i),
\end{equation}
where the angular samples $\varphi_j=2\pi j/M$ are equally spaced and $M$ is chosen to match the angular speckle scale at $r=q_\mathrm{max}$. From the data reduction stage we know these harmonics up to an overall phase:
\begin{equation}
I_{mi}'=e^{i \alpha_m}I_m(r_i,z_i)=e^{i \alpha_m}\tilde{I}_{m i}.
\end{equation}
All that remains is to determine the angles $\alpha_m$ by minimizing
\begin{equation}\label{metric2}
\sum_{m=0}^{M-1} \sum_i|I_{m i}'-I_{m i}|^2=\sum_{m=0}^{M-1} \sum_i|e^{i \alpha_m}\tilde{I}_{m i}-I_{m i}|^2.
\end{equation}
The metric of \eqref{metric2} is equivalent to the metric \eqref{metric1} by Parseval's theorem. Since $\alpha_0=0$, the projection simply replaces $I_{0 i}$ by the powder average $\tilde{I}_{0 i}$. For $m\ne 0$ the minimization of \eqref{metric2} yields
\begin{subequations}
\begin{align}
I_{m i}&\rightarrow e^{i \alpha_m}\tilde{I}_{m i},\\
\alpha_m&=\arg{\left(\sum_i \tilde{I}_{m i}^\ast I_{m i}\right)}.
\end{align}
\end{subequations}

So far we have defined two projections
\begin{equation}
P_D, P_C\colon \quad\{\hat{\rho},I\}\rightarrow \{\hat{\rho}',I'\},
\end{equation}
where $P_D$, the ``divided constraint projection", imposes the support constraint \eqref{con1} on $\hat{\rho}$ and the data constraint \eqref{con2} on $I$, while the ``concurrence projection" $P_C$ imposes the compatibility constraint \eqref{con3} between $\hat{\rho}$ and $I$ (while ignoring the support and data constraints). As distance minimizing operations, these are in a sense ``greedy" moves to their corresponding constraint sets. Solutions $\{\hat{\rho}_\mathrm{sol},I_\mathrm{sol}\}$ to the reconstruction problem have the property that they are fixed by both projections, and in particular, $\hat{\rho}_\mathrm{sol}$ is simultaneously consistent with the intensity correlation data and a compact support. There are provably convergent algorithms for finding fixed points in a general projection setting when the constraint sets are convex. Because in our case constraints \eqref{con2} and \eqref{con3} are non-convex, convergence cannot be proven and we are forced to use algorithms that have a demonstrated record of discovering solutions even for difficult non-convex problems. We use the difference map iteration  \cite{diffMap}
\begin{equation}
\{\hat{\rho},I\}\rightarrow \{\hat{\rho},I\}+P_C\left(2 P_D\{\hat{\rho},I\}-\{\hat{\rho},I\}\right)-P_D\{\hat{\rho},I\},
\end{equation}
which is equivalent to Fienup's hybrid input-output iteration \cite{Fienup} in the standard reconstruction problem with support and Fourier magnitude as the two constraints ($P_C\to P_S$, $P_D\to P_F$). For our initial iterates we generate a random positive density $\rho$ (uniformly distributed contrast) on the support and from this obtain $\hat{\rho}$ and $I=|\hat{\rho}|^2$, a configuration that only satisfies the compatibility constraint. Iterations are performed until the metric $\Delta$ of the updates \eqref{metric1} fluctuates about a steady state and shows no sign of further decrease.

\subsection{Reconstruction of axial projections}\label{reconaxial}\label{axialrecon}

\begin{figure}[t]
\centering
\includegraphics[width=5.in]{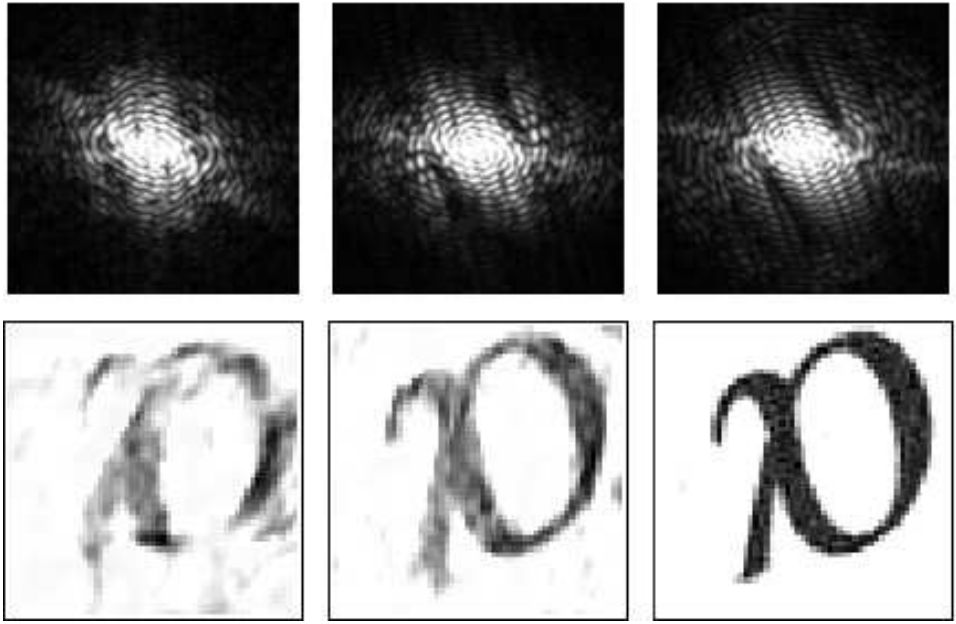}
\parbox{6in}{\caption{Iterations 40, 70 and 200 in the simultaneous reconstruction of intensity (top) and density (bottom) of the test particle axial projection (the greek letter $\alpha$) using the algorithm described in section \ref{axialrecon}. The square frames in the bottom row are the boundary of the support.}}
\end{figure}

The specialization of the reconstruction algorithm to axial projections amounts to simply setting $z=0$ in the equations of the previous section. In the flat Ewald sphere limit we can also take advantage of Friedel symmetry, thereby reducing by a factor of 2 the work in projecting to the data constraint (the angular Fourier transforms act on a periodic function on the domain $0\le\varphi<\pi$).

The advantage of simultaneously reconstructing intensity and particle density is evident from the frame by frame development of these in the early iterations of the difference map. There is a ``locking" of the angles $\alpha_m$, that determine the intensity reconstruction, at about the same point in time that the density is well localized in the support, a necessary condition to have speckles in the intensity. In the subsequent refinement iterations the particle density and intensity often executed significant (synchronized) rotation. We deliberately chose a rather loose support to enable this freedom in the reconstruction. Figure 6 shows three frames in a reconstruction with low noise ($\mathrm{SN}=10^4$). The difference map error metric is plotted in Figure 7 for three levels of noise and not surprisingly shows that the fixed-point property ($\Delta\to 0$) is strongly compromised when the noise is high. The corresponding reconstructions are shown in Figure 8.

\begin{figure}[t]
\centering
\includegraphics[width=5.in]{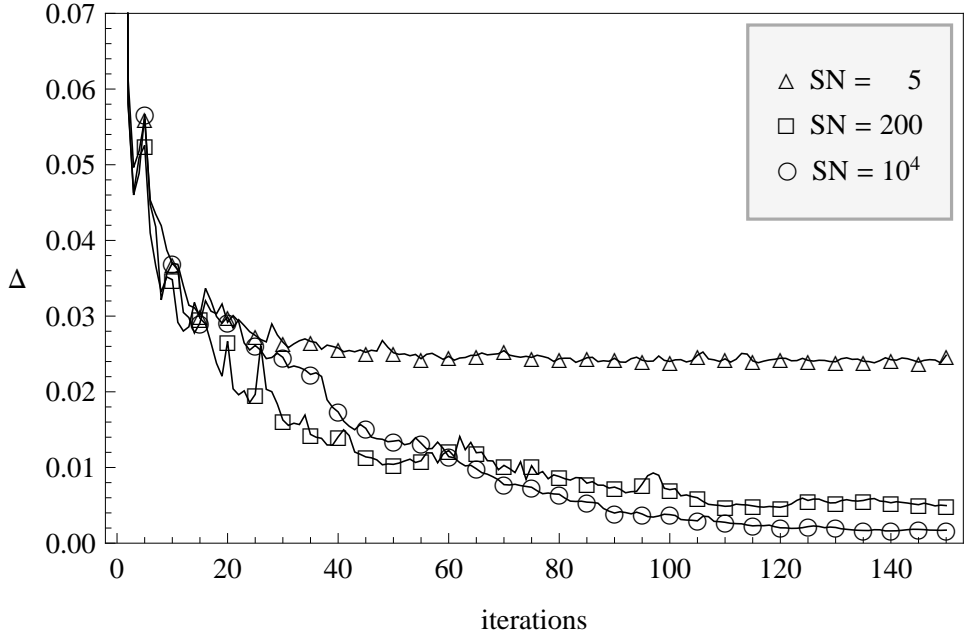}
\parbox{6in}{\caption{Time series of the difference map error metric $\Delta$ in test particle axial projection reconstructions for three values of noise. The corresponding particle densities are shown in Figure 8.}}
\end{figure}

\begin{figure}[b]
\centering
\includegraphics[width=5.in]{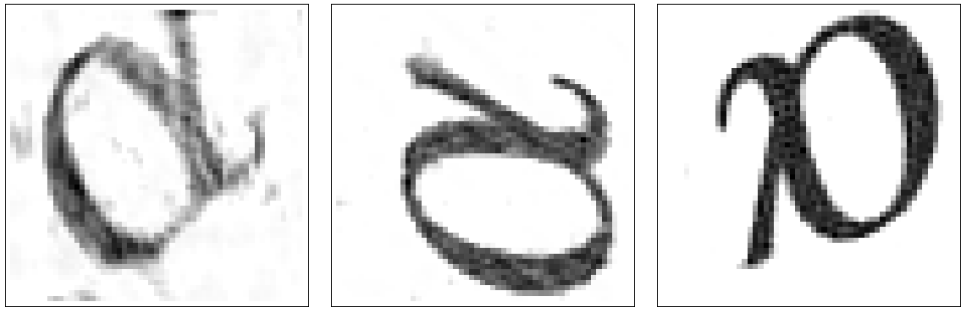}
\parbox{6in}{\caption{Reconstructions of the test particle at SN values 5, 200, and $10^4$.}}
\end{figure}

Given the relatively small number of phase angles required for intensity reconstruction, the weaker constraint arising from the support of the intensity Fourier transform --- the density autocorrelation --- might avoid the complications in the simultaneous intensity/density reconstruction described above. The latter constraint is a weaker constraint on the intensity in that the Fourier transform is only required to have a particular support, and not necessarily the structure of an actual autocorrelation. 
We tested this simplification for our test particle by pairing the projection to the data constraint \eqref{con2} with a simple autocorrelation support projection in the difference map scheme. This completely avoids manipulating the particle density and the non-linear compatibility constraint \eqref{con3}. Whereas this simpler approach did succeed in reconstructing the intensity, the progress of the algorithm, shown in Figure 9, was much slower and more sensitive to noise. And since the density is always the final goal of the reconstruction anyway, this two-stage approach does not appear to offer any advantages over the simultaneous intensity/density reconstruction.

\begin{figure}[t]
\centering
\includegraphics[width=5.in]{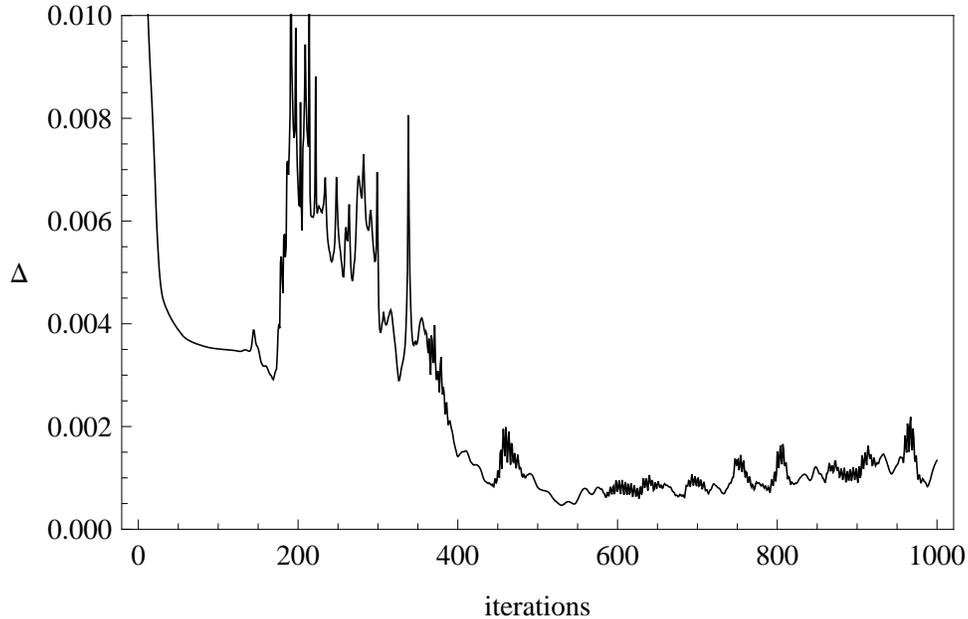}
\parbox{6in}{\caption{Difference map error metric when reconstructing just the test particle intensity from low-noise data and the autocorrelation support constraint. Convergence is much slower with this weaker prior constraint than for the joint intensity/density reconstructions (Fig. 7). A well reconstructed intensity (not shown) first appears only after 500 iterations.}}
\end{figure}

\section*{Acknowledgements}

I thank John Spence for suggesting this problem and Duane Loh for help with the figures. This work was supported by Department of Energy Grant DE-FG02-11ER16210.

\end{document}